\documentclass[aip,jcp,preprint]{revtex4-1}
\usepackage{graphicx}
\usepackage{amssymb}
\usepackage{amsmath}
\usepackage{chemfig}
\usepackage{stmaryrd} % [[ and ]]
\usepackage{subcaption}
\DeclareRobustCommand*{\citen}[1]{%
  \begingroup
    \romannumeral-`\x % remove space at the beginning of \setcitestyle
    \setcitestyle{numbers}%
    \cite{#1}%
  \endgroup
}

\newlength{\figwidth}
\newcommand{\fref}[1]{Fig.\,\ref{#1}}

\newcommand{\eref}[1]{Eq.\,(\ref{#1})}

\newcommand{\sref}[1]{Sec.\!~\ref{#1}}

\newcommand{\cref}[1]{Ref.\,\citen{#1}}
\newcommand{\crefs}[1]{Refs.\,\citen{#1}}
\newcommand{\eg}{{\it e.g.}\ }

\newcommand{\etal}{{\it et al.}\ }

\newcommand{\abinitio}{{\it ab initio} }

\newcommand{\eb}{\mathbf{e}}

\newcommand{\xb}{\mathbf{x}}

\newcommand{\Eb}{\mathbf{E}}

\newcommand{\grad}{\boldsymbol{\nabla}}

\begin{document}
\title{\bf Insight into hydrogen production through molecular simulation of an electrode-ionomer electrolyte system}
\author{R.E. Jones}\email{rjones@sandia.gov} 
\affiliation{Sandia National Laboratories, Livermore, CA 94551, USA}
\author{W.C. Tucker}
\affiliation{Sandia National Laboratories, Livermore, CA 94551, USA}
\author{M.J.L. Mills}
\affiliation{Sandia National Laboratories, Livermore, CA 94551, USA}
\author{S. Mukerjee}
\affiliation{Northeastern University, Boston, MA 02115, USA}

\date{\today}



\begin{abstract}
In this work, we examine metal electrode-ionomer electrolyte systems at high voltage / negative surface charge and at high pH to assess factors that influence hydrogen production efficiency.
We simulate the hydrogen evolution electrode interface investigated experimentally in Bates {\it et al.}, {\it Journal of Physical Chemistry C}, 2015 using a combination of first principles calculations and classical molecular dynamics.
With this detailed molecular information, we explore the hypotheses posed in Bates \etal
In particular we examine the response of the system to increased bias voltage and oxide coverage in terms of the potential profile, changes in solvation and species concentrations away from the electrode,  surface concentrations, and orientation of water at reactive surface sites.
We discuss this response in the context of hydrogen production.
\end{abstract}

\maketitle

\section{Introduction}

Hydrogen has the capacity to be an ecologically-friendly fuel since water is the primary by-product of its use.
Many technological and economic challenges remain in realizing a viable hydrogen economy and energy system \cite{rahman2006special,nrc2004hydrogen}.
The central issue is that molecular hydrogen gas does not occur naturally in abundance and must be produced industrially.
Currently, the majority of hydrogen is generated by either high-temperature/high-energy  methane reforming or the water-gas shift reaction which produces significant carbon dioxide, while the electrolysis of water accounts for a relatively minor proportion of its production \cite{navlani2018recent}.
As electrolysis is not directly dependent on fossil fuels there is strong motivation to develop a low cost, low energy electrolytic process.
Since electrolysis traditionally depends on expensive Pt-group catalysts, transition metal catalysts are being developed.  

Mukerjee and co-workers have been developing Ni based catalysts with ionomer-based electrolytes \cite{bates2015composite,unlu2011analysis} which show promise but are still confronted by development challenges in part due to the need for a more fundamental understanding of the electrode-electrolyte interaction.
In Bates \etal \cite{bates2015composite} a number of hypotheses were put forward:
\begin{itemize}
\item The electrical potential is significantly altered by the ionomer. 
The ionomer extends the ``electrified interface'' near the catalyst.
The potential at the inner Helmholtz plane dictates the rate of the hydrogen evolution reaction (HER).
This potential is more positive than it would be in absence of ionomer and this affects the reactant water molecules at the inner plane.
\item The water molecules inside the inner Helmholtz plane orient with dipoles pointing away from the electrode due to the presence of the ionomer, and this facilitates H–OH cleavage on the metal catalyst
\item The ionomer itself straddles the inner and outer Helmholtz planes and is directly chemisorbed on the metal surface.
\item The majority of the metal surface is coordinated with water molecules.
\item Nanoscale heterogeneity provides a high density of adjacent metal/metal-oxide sites where metallic Ni has an affinity for H-bonding, and NiO$_x$ has an affinity for adsorbed OH$^-$, in-line with Markovic's theory of enhanced hydrogen evolution reaction on composite metal/metal-oxide surfaces \cite{subbaraman2011enhancing,strmcnik2013improving,danilovic2012enhancing}.
\end{itemize}

In this work we investigate these hypotheses via molecular simulation.
Since a system encompassing the compact/diffuse layers of the long chain ionomer-based electrolyte is too large for \abinitio calculations such as \crefs{filhol2006elucidation,jinnouchi2008electronic,skulason2010modeling,skulason2007density,karlberg2007cyclic,taylor2006first,rossmeisl2006calculated,janik2009first,otani2008electrode}, which allow charge transfer and spontaneous dissociation, we employ classical molecular dynamics (MD) to model the system at the relevant length and time scales.
Water-metal interfaces have been studied with MD for some time \cite{spohr1999molecular,willard2009water,limmer2013hydration,yeh2013molecular,hughes2014structure,dewan2014structure,limmer2015water,takae2015fluctuations,foroutan2017structural}.
The absence of chemical reactions in these simulations is offset by the fact that these reactions are fast compared to diffusion timescales resolvable with MD and we pre-populate the simulation with the relevant chemical species.
We assume simulations of the isothermal steady-state accounting for electrostatic and steric interactions  with the experimentally observed species concentrations is informative of the transport-limited steady operation of the actual cell.
Also, we apply \abinitio methods to assess the near electrode charge environment and use this information in the MD model of the electrode.
We focus on the hydrogen evolution reaction (HER) environment at the negatively charged electrode (cathode).
Here the ionomer is in contact with bare metal (Ni) and metal oxide (NiO$_x$) at the electrode surface.
The ionomer-based electrolyte has characteristics that differ from the well-studied dissolved salt electrolytes.
For instance, the ionomer is relatively immobile due to its polymeric structure with fixed charge centers (N$^+$), while the counter ions (OH$^-$) are mobile.
The influence of the immobile charge in ionomer strands/chains on performance is  central to our investigation.

Given its importance, the structure and chemistry of water near a metal surface has long and intense field of study in and of itself, which is reviewed in \crefs{thiel1987interaction,henderson2002interaction,hodgson2009water,carrasco2012molecular,stuve2012ionization}.
The examination of the water-surface interactions has been pursued in great detail by combinations of microscopy and \abinitio simulation (typically in vacuum), as in \cref{nie2010pentagons}, and has lead to many postulated and observed intact, partially dissociated, and fully dissociated configurations of water at uncharged atomically flat and rough surfaces. 
Despite the limitations of classical MD, we attempt to interpret the results of this study in the context of this deep body of research.

In \sref{sec:theory} we briefly review the relevant theory.
With representative electrode-ionomer systems we simulate the response of these systems to a range of external electrical bias and characterize this response by a variety of means to address the hypotheses in Bates \etal \cite{bates2015composite}.
In \sref{sec:method} we describe how we use spatial resolution of charge density, per-species radial distributions and other fields to quantify the location of the Helmholtz planes and significant concentrations of the reactive species.
The results are given in \sref{sec:results} and are discussed in \sref{sec:discussion} in light of the hypotheses of Bates \etal \cite{bates2015composite}

\section{Theory} \label{sec:theory}

Due to high electric fields near electrodes, charged mobile species tend to pack and form characteristic structures and concentration profiles.
This phenomena is modelled in classical theory by Helmholtz \cite{helmholtz1853ueber}, Gouy  \cite{gouy1910constitution}, Chapman \cite{chapman1913li}, Stern \cite{stern1924theorie} and others, and is still a topic of current research \cite{merlet2014electric}.
Here we will briefly review the relevant theory to assist in interpreting the molecular results in \sref{sec:results}.

The most commonly used theory is the Gouy-Champman-Stern (GCS) model, composed of a compact layer of ions next to the electrode and a diffuse layer beyond.
The compact layer is bounded by the inner Helmholtz plane (IHP) at a layer of unsolvated ions adsorbed on the electrode surface and the outer Helmholtz plane (OHP) where ions are fully solvated. 
The interior of the compact layer is assumed to be charge free due to steric effects.
The structure of the diffuse layer is governed by the Poisson-Boltzmann (PB) equation where the electrostatic interaction of the ions is given by the Poisson equation:
\begin{equation} \label{eq:Poisson}
-\grad \cdot (\epsilon \grad \phi ) = \rho \ .
\end{equation}
Here, $\phi$ is the electric potential and $\epsilon$ is the electric permittivity.
The net charge density $\rho$ is given by:
\begin{equation} \label{eq:rho}
\rho = e \sum_a z_a c_a \ ,
\end{equation}
where $z_a$ is the valence of species $a$, $c_a$ is its concentration, and $e$ is the unit of elemental charge.
In equilibrium, the absence of fluxes implies that both the electrochemical potential, $e z_a  \phi + k_B T \ln c_a$, and the temperature, $T$, are constant across the system domain.
This condition leads to species concentrations, $c_a$, that vary with the local potential $\phi$:
\begin{equation} \label{eq:Boltzmann}
c_a = \bar{c}_a \exp\left(\frac{e z_a \phi}{k_B T}\right) \ ,
\end{equation}
where $\bar{c}_a$ is the far-field/bulk concentration of species $a$.
Substituting \eref{eq:rho}, and \eref{eq:Boltzmann} in \eref{eq:Poisson} results in the Poisson-Boltzmann (PB) equation.
The characteristic thickness of the diffuse layer is given by the Debye length 
\begin{equation}
 \lambda_D = \sqrt{\frac{k_B T \epsilon}{e^2 \sum_a z_a^2 \bar{c}_a}} \ ,
\end{equation}
which is the similarity parameter in the solution of the linearized PB equations.
For more details see \cref{BardFaulkner}(Sec.13.3), \cref{Israelachvili} (Chap.12), and \cref{Newman} (Sec.7.4).

Real electrode-electrolyte systems deviate from GCS for a number of reasons such as: finite ion size and correlation effects due to van der Waals interactions, and a non-uniform dielectric field due to varying concentration and discrete charges \cite{kornyshev2007double,oldham2008gouy,bazant2011double}.
Also theoretical concepts such as full solvation and the location of Helmholtz planes become less well-defined.
In the present case, an electrolyte with a dense ionomer, the charge centers of ionomer chains are effectively immobile and a simple model assumes these ionomer charges provide a background charge density which is fully solvated and screened far from the electrode.
It follows that the PB equation governs the excess charge of the counter ion.
This paradigm holds where counter ions stay bound to ionomer charge centers, and fails where the electric field is strong enough to dissociate the mobile species from the charge centers. 
The solution to the PB equation for the counter ions only \cite{gray2018nonlinear} is:
\begin{equation} \label{eq:PB}
\phi(x) = -2 \frac{k_B T}{e} \ln\left(1 + \frac{x}{\ell}\right) + \phi_0 
\end{equation}
and 
\begin{equation}
c(x) = \frac{c_0}{\left(1 + \frac{x}{\ell}\right)^2} \ ,
\end{equation}
where $\phi_0$ is the potential at $x=0$, $\ell = 2 \frac{\epsilon k_B T}{e \sigma}$ and $\sigma$ is the surface charge of the electrode.
Note that the dielectric of the electrolyte, $\epsilon$, is assumed to be spatially uniform in this derivation.
This diffuse layer solution has an electric field $\Eb = - \grad \phi$ that is zero far away from the electrode and has the magnitude $\frac{\sigma}{\epsilon}$ at $x=0$, which is consistent with Gauss's law applied to the electrode (this discussed in more detail in the next section).

\section{Method} \label{sec:method}

To simulate the metal electrode-ionomer based electrolyte system, we employ a combination of density functional theory (DFT) and classical molecular dynamics (MD).
We use DFT to determine the charge state of the metal HER electrode, which is our primary focus.
We use MD to model the dynamics of the molecular species of the electrolyte through interplay of electrostatic, elastic, steric and diffusion forces in the overall system.
By its explicit representation of atoms and atomic bonding, MD is known to capture deviations from classical theories such as GCS \cite{lee2012comparison,lee2013comparison,lee2015atomistic}.

In the molecular model, the electrode-electrolyte system atoms and molecules interact with each other via the well-known CHARMM empirical potential \cite{mackerell2001atomistic}.
This often employed potential depends on atomic charges and proximity.
It is composed of short-range van der Waals  interactions,  long-range Coulomb interactions, and intramolecular covalent bonds:
\begin{align} \label{eq:md_potential}
\Phi(\xb_\alpha) 
&=  \underbrace{\sum_{\alpha <\beta} 4\varepsilon_{ab} \left( \left(\frac{\sigma_{ab}}{r_{\alpha\beta}} \right)^{12} - \left(\frac{\sigma_{ab}}{r_{\alpha\beta}} \right)^6  \right)}_\text{van der Waals}
+ \underbrace{\frac{1}{4 \pi \epsilon_0} \sum_{\alpha <\beta} \frac{q_\alpha q_\beta}{r_{\alpha\beta}}}_\text{Coulomb}  \\
&+ \underbrace{ \sum_I \left[ \sum_{\alpha,\beta\in M_I}  k_{ab} r^2_{\alpha\beta}
+ \sum_{\alpha,\beta,\gamma\in M_I}  k_{abc} \theta^2_{\alpha\beta\gamma}
+ \sum_{\alpha,\beta,\gamma,\mu\in M_I}  k_{abcd} \phi^2_{\alpha\beta\gamma\mu}
\right]}_\text{covalent} \nonumber
\end{align}
where $\varepsilon_{ab}$ and $\sigma_{ab}$ are the usual Lennard-Jones (LJ) pair parameters for species $a$ and $b$,  $q_\alpha$ is the charge of atom $\alpha$, $r_{\alpha\beta} = \| \xb_\alpha - \xb_\beta\|$ is the distance between atoms $\alpha$ and $\beta$, and $a,b,c,d$ index atom types.
We employ a particle-particle particle-mesh (PPPM) solver \cite{hockney1988computer} to efficiently compute the long-range Coulomb interactions with a short/long cutoff of 12 \AA\ .
Here, the permittivity of free space is $\varepsilon_0$= 0.00552635 e/V-\AA\ .
The intra-molecular bonds are effected by harmonic potentials based on pair distances $r_{\alpha\beta}$, 3 atom angles $\theta_{\alpha\beta\gamma}$, and 4 atom dihedral angles $\phi_{\alpha\beta\gamma\mu}$, where $M_I$ is a set of like molecules.

\subsection{Electrode} \label{sec:electrode}
Given the form of the interatomic potential, \eref{eq:md_potential}, which includes Coulomb forces,  we need the point charges $q_\alpha$ for the classical representation of the electrode via \eref{eq:md_potential} and the response of the electrode point charges to electrical bias/external potential $V$.
We use DFT with PBE/GGA level of theory \cite{perdew1996generalized} to obtain relaxed surface structures, and compute point charges via a Bader analysis of the charge density field \cite{tang2009grid}. 

Since only studies of small systems are feasible with DFT, we examined specific domains of a partially oxidized Ni electrode: bare Ni, partially oxidized Ni, and Ni covered by an NiO layer.
For each, we create a small, laterally periodic system to calculate charge density using a FCC unit cell with a 3.52 \AA\ lattice constant for Ni regions and a cubic B1 unit cell with a 4.17 \AA\ lattice constant for NiO regions.
We select (100)-oriented (non-polar) surfaces for both Ni and NiO.
The surfaces neighbor vacuum regions, not representative electrolytes, for simplicity and under the assumption that the proximity of the electrolyte evokes only perturbations of the charge density.
We employ an energy cutoff of 500 eV for the metallic systems and 800 eV for the systems containing oxides, together with an 8$\times$8$\times$1 $\Gamma$ centered $k$-point grid.
After computing a baseline charge density, we add excess electrons to emulate negative charging of the electrode of interest.
Finally, we obtain the electrode point charges $q'_\alpha(\sigma)$:
\begin{equation}\label{eq:qperturbed}
q'_\alpha(\sigma) = q_\alpha - \frac{\sigma}{e \eta} \, \Delta q_\alpha
\end{equation}
where $q_\alpha$ are the baseline point charges (such that $\sum_\alpha q_\alpha = 0$), $\Delta q_\alpha$ are the perturbed point charges (corrected for the homogeneous background charge and normalized such that $\Delta q_\alpha$ = 1 $e$ for the surface atoms), $\sigma$ is the target surface charge density, and $\eta$ is surface atom density. 
The number of excess electrons added the systems to obtain the perturbed charge field is on the order of 0.1 $e$ per surface atom and on par with the perturbation needed to achieve the voltage bias $V$ in the range $V \approx$ 0--2.5 V in the experimental system.
Lastly, the scaling of \eref{eq:qperturbed} with external voltage $V$ follows $V\approx \frac{\lambda_D}{\varepsilon_0} \sigma$; however, we recover the actual voltage in the MD systems via a Gauss box method described in \sref{sec:method}.

In addition to the point charges required for Coulomb interactions, short-range parameters $\varepsilon_{ab}$ and $\sigma_{ab}$ for the electrode interactions with the ionomer are needed.
By assuming traditional Lorentz-Berthelot mixing, only the LJ self-self pair parameters are required.
We obtain these from published, surface specific parameterizations: for Ni, 
$\sigma_\text{NiNi} = $ 2.274 \AA\ and $\epsilon_\text{NiNi}$= 5.65 kcal/mol
\cref{heinz2008accurate}(Table 1):
and for NiO,  
$\sigma_\text{OO}   = $ 1.292 \AA\ and $\epsilon_\text{OO}$=  35.62 kcal/mol 
\cref{oliver1995molecular}(Table 1).%
\footnote{\cref{oliver1995molecular} reports parameters for a Buckingham potential which we convert to a LJ parameterization by matching the potential well location and stiffness.}
For simplicity (since no additional parameters are needed) and efficiency we neglect thermal motion of the electrode and fix the locations of its Ni and O atoms.

\subsection{Electrolyte} \label{sec:electrolyte}
\begin{figure}[htp!]
\centering
{\includegraphics[width=2.0\figwidth]{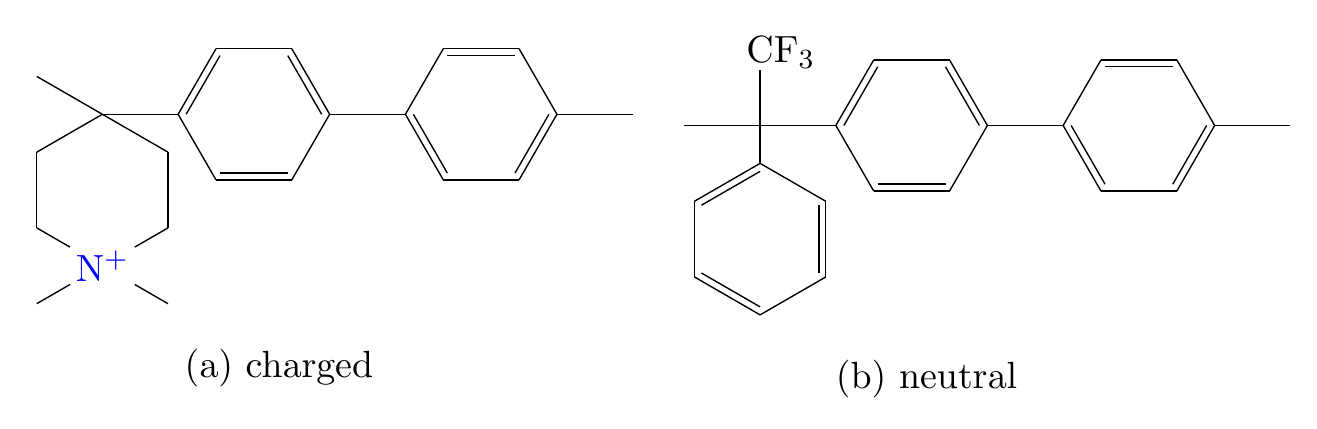}}
%\begin{subfigure}[c]{0.3\textwidth}
%\centering
%{\setchemfig{atom sep=2.0em}\chemfig{
%-[:-30]([:-90]*6(---\textcolor{blue}{N^+}(-[:-150])(-[:-30])----))
%-[:0]*6(-=-(-*6(=-=(-)-=-))=-=)
%}}
%\caption{charged}
%\end{subfigure}\hspace{0.1\textwidth}%\hfill%
%\begin{subfigure}[c]{0.3\textwidth}
%\centering
%{\setchemfig{atom sep=2.0em}\chemfig{
%-[:0](-[:-90,0.5]*6(=-=-=-))(-[:90]CF_3)
%-[:0]*6(-=-(-*6(=-=(-)-=-))=-=)
%}}
%\caption{neutral}
%\end{subfigure}\hfill
\caption{PAP-DP-60 ionomer units: (a) charged (60\%), and (b) neutral units (40\%).
The ionomer is charge balanced by OH$^-$.}
\label{fig:units}
\end{figure}

The selected ionomer (PAP-DP-60) is an amorphous material composed of long polymer chains with N$^+$ charge centers charge-balanced by OH$^-$ together with water.
Each chain is comprised of charged (c) and neutral units (n), see \fref{fig:units}, in a ratio c:n = 60:40 with lengths of 30 to 40 units in a random sequence.
The mass density of the ionomer is 1.1 g/cm$^3$ with about 40\% water by weight at room temperature and pH in the range 13 to 14+.

\begin{figure}[htp!]
\centering
{\includegraphics[width=2.0\figwidth]{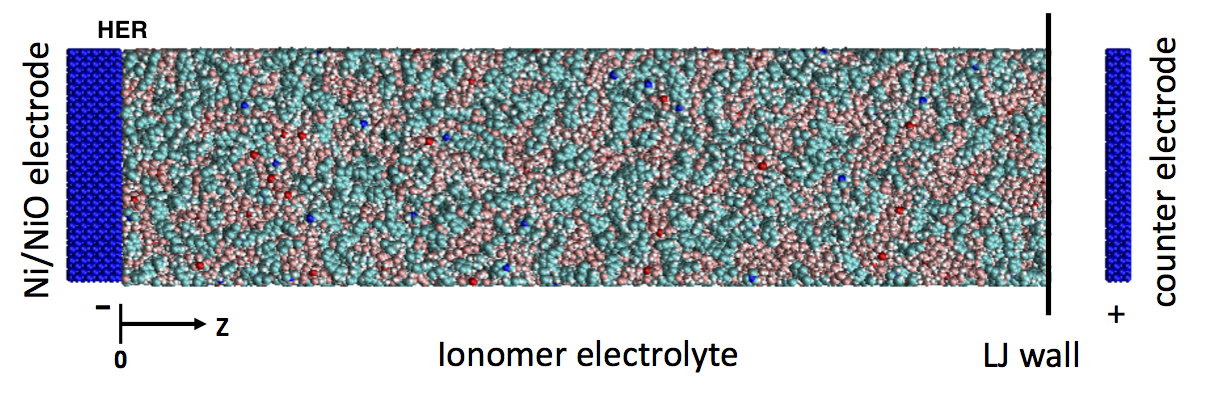}}
\caption{Electrode:electrolyte system. 
The ionomer membrane infused with OH$^-$ counter ions and water.
The metal electrode has imposed surface charges to effect electrical bias.
The replica systems have dimensions: 6-7 nm $\times$ 6-7 nm $\times$ 34-38 nm each consisting of 32 unique ionomer chains with random sequences.
The origin $z=0$ is a plane through centers of the surface layer atoms of the Ni/NiO (HER) electrode.
}
\label{fig:system}
\end{figure}

Using these experimental measurements, we created representative models of the ionomer electrolyte, see \fref{fig:system}.
First we created chains from the units using a random sequence that respected the experimental bounds on length and c:n ratio.
Next we added molecular water and OH$^-$ ions.
To achieve the experimentally measured density, we compressed the ionomer-electrolyte mixture then relaxed the compressed configuration at 2000 K via isothermal dynamics to relieve unphysical local configurations for 0.4 ns.
Finally, we cooled the system to 300 K at 10 K/ps and let it equilibrate for 0.1 ns.
To complete the electrode-electrolyte system seen in \fref{fig:system}, we added the primary Ni electrode plus a soft wall and counter electrode to bound the system.
The separation of the capping wall and the counter electrode was expedient due to different densities and lengths of the replicas while maintaining the same effective gradient in external potential.
Due to computational cost we can only place the counter electrode 400 nm from the surface of the primary electrode.
As stated in the introduction, our focus is on the Ni electrode where the HER occurs and, hence, we model it in detail and simplify the counter electrode.
For our purposes this is sufficient as long as the diffuse layer near the primary electrode relaxes to the bulk, and the region away from the primary electrode is not depleted of mobile species.
As we will see in \sref{sec:results} both these conditions are satisfied.
Also, in preliminary studies, the lateral dimensions of the systems where increased to the point where the charge density profiles and related measures converged to within the expected statistical noise.

Since the ionomer is amorphous, we generated a number of statistically equivalent replica systems to improve sampling and reduce finite size effects.
The isothermal dynamics of the electrolyte were evolved with a Nos\'e-Hoover algorithm which accommodated the rigidity of the intramolecular bonds.
All reported results are the average of 8 configurations time-averaged and simulated for 0.4 ns each with a 1 fs time-step.


Finally, we use a so-called Gauss box to recover the electric potential $\phi$ from the point charges $q_\alpha$. 
Briefly, starting with Gauss's law for a quasi-one dimensional system, with electric field $\Eb \equiv -\grad \phi = E \eb_3$ and cross-sectional area $A$, we obtain:
\begin{equation}
\phi(z)    = 
\phi(0)
-\frac{1}{\varepsilon_0 A} \int_0^z \overbrace{\bigl< \sum_{\alpha | z_\alpha\in[0,\tilde{z}]} q_\alpha \bigr> }^{Q(\tilde{z})} \,  \mathrm{d}\tilde{z}  \ ,
\end{equation}
where $Q(\tilde{z})$ is the total/net charge in $z\in[0,\tilde{z}]$ and $\left< \bullet \right>$ is the average over replica systems and steady-state trajectories in each system.
In a similar fashion, we estimate charge density, dipole density, and species concentration profiles using coarse-grained point charges $q_\alpha$, atomic dipoles $q_\alpha \xb_\alpha$, and species type.
For instance, the water dipole density is given by:
\begin{equation} \label{eq:dipole}
\mu(z_I) = \frac{1}{V_I} \Bigl< \sum_\alpha  q_\alpha z_\alpha w(z_I-z_\alpha) \Bigr> - \frac{1}{V_I} q_I z_I \ ,
\end{equation}
where $z_I$ is the location of points on the coarse-grained sampling grid,
$V_I$ is the effective volume associated with $I$,
$w$ is a partition-of-unity, tent-like kernel function, and 
\begin{equation}
q_I =  \Bigl< \sum_\alpha  q_\alpha w(z_I-z_\alpha) \Bigr> \ .
\end{equation}
In  \eref{eq:dipole}, the second term is a correction for net charge in a particular bin $I$.
Also, we use a kernel width of 3 \AA\ based on the size of water molecules, and cutoff the averaging kernel $w(z)$ at electrode surface so that the average only includes the time-dependent data in the electrode.
The water dipole density is used to determine water orientation and the species concentrations indicate which species are present in the compact and diffuse layers.
Also, we extract coarse-grained profiles of radial distribution functions (RDFs) of particular species to quantify changes in solvation with 3 \AA\ resolution.
This information allows us to estimate the location of the OHP and its response to bias.

\section{Results} \label{sec:results}

\begin{figure}[htp!]
\centering
\begin{subfigure}[b]{0.30\textwidth}
\centering
{\includegraphics[width=1.0\textwidth]{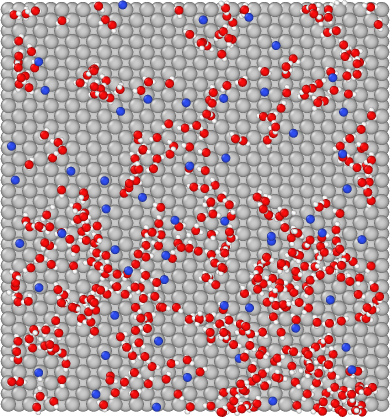}}

{\includegraphics[width=1.0\textwidth]{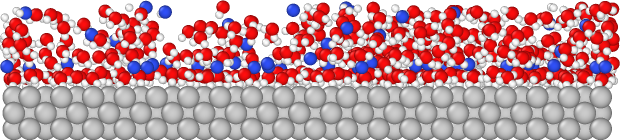}}

\caption{bare Ni}
\end{subfigure}
\begin{subfigure}[b]{0.30\textwidth}
\centering
{\includegraphics[width=1.0\textwidth]{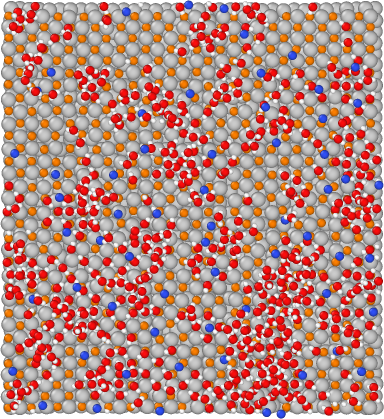}}

{\includegraphics[width=1.0\textwidth]{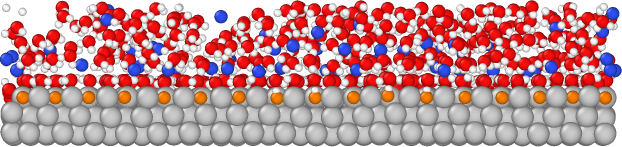}}

\caption{NiO monolayer}
\end{subfigure}
\begin{subfigure}[b]{0.30\textwidth}
\centering
{\includegraphics[width=1.0\textwidth]{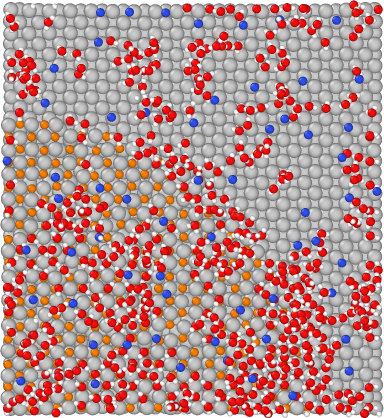}}

{\includegraphics[width=1.0\textwidth]{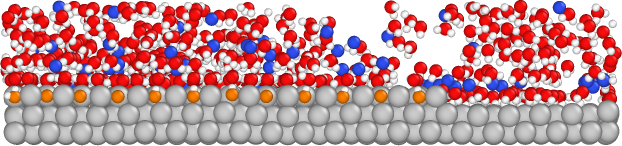}}

\caption{partial NiO coverage}
\end{subfigure}
\caption{Surface coverage of H$_2$O, OH$^-$, and N$^+$ in the ionomer for $q$=-0.10 $e$/atom (top-down and side views of instantaneous configurations).
(a) bare Ni,
(b) Ni covered by a monolayer of NiO,
(c) Ni with $\approx$50\% partial NiO coverage quarter disk monolayer centered on lower left corner of top view.
Atom color map, H:white, O:red, N$^+$:blue, Ni:gray, O in NiO:orange.
}
\label{fig:surface_species}
\end{figure}

\begin{figure}[htp!]
\centering
\begin{subfigure}[b]{0.30\textwidth}
\centering

{\includegraphics[width=0.8\textwidth]{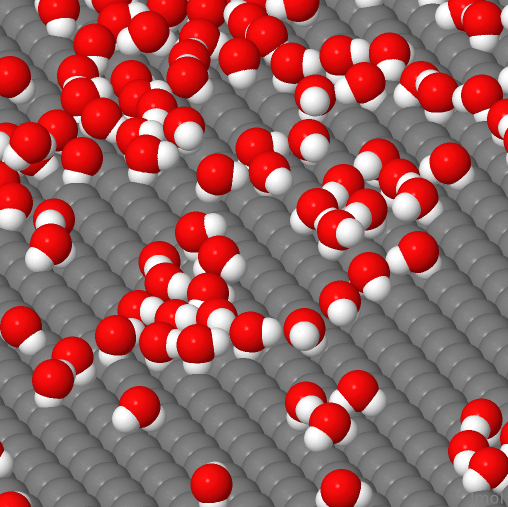}}

\caption{bare Ni}
\end{subfigure}
\begin{subfigure}[b]{0.30\textwidth}
\centering

{\includegraphics[width=0.8\textwidth]{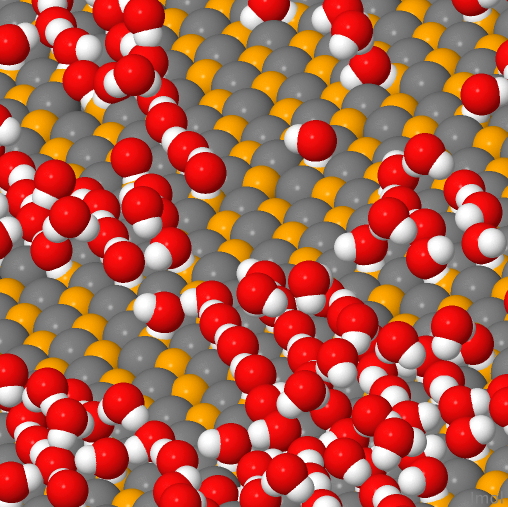}}
\caption{NiO monolayer}
\end{subfigure}
\begin{subfigure}[b]{0.30\textwidth}
\centering

{\includegraphics[width=0.8\textwidth]{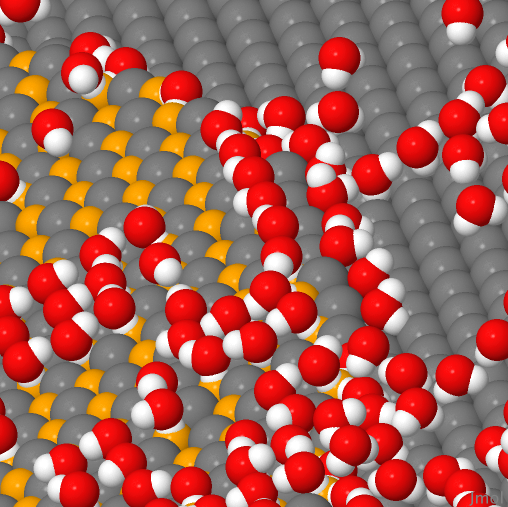}}
\caption{partial NiO coverage}
\end{subfigure}
\caption{Surface orientation of H$_2$O, for $q$=-0.10 $e$/atom (instantaneous configurations).
(a) bare Ni,
(b) Ni covered by a monolayer of NiO,
(c) Ni with $\approx$50\% partial NiO coverage quarter disk monolayer centered on lower left corner of top view.
Atom color map, H:white, O:red, N$^+$:blue, Ni:gray, O in NiO:orange.
}
\label{fig:surface_orientation}
\end{figure}

We compare the response of the electrolyte to electrode bias voltage using a variety of configurational metrics in order to provide insight and support to the experimental hypotheses described in the Introduction.
We focus on (a) bare Ni, (b) fully NiO covered and (c) partially NiO covered regions of the Ni electrode.
\fref{fig:surface_species} and \fref{fig:surface_orientation} show the qualitative differences in surface coverage of the relevant species and water orientation for the three cases.
\fref{fig:surface_species} and \fref{fig:surface_orientation} will be discussed in more detail in the following sections.

\subsection{Electrode charge distributions}
\begin{figure}[htp!]
\centering
{\includegraphics[width=0.7\figwidth]{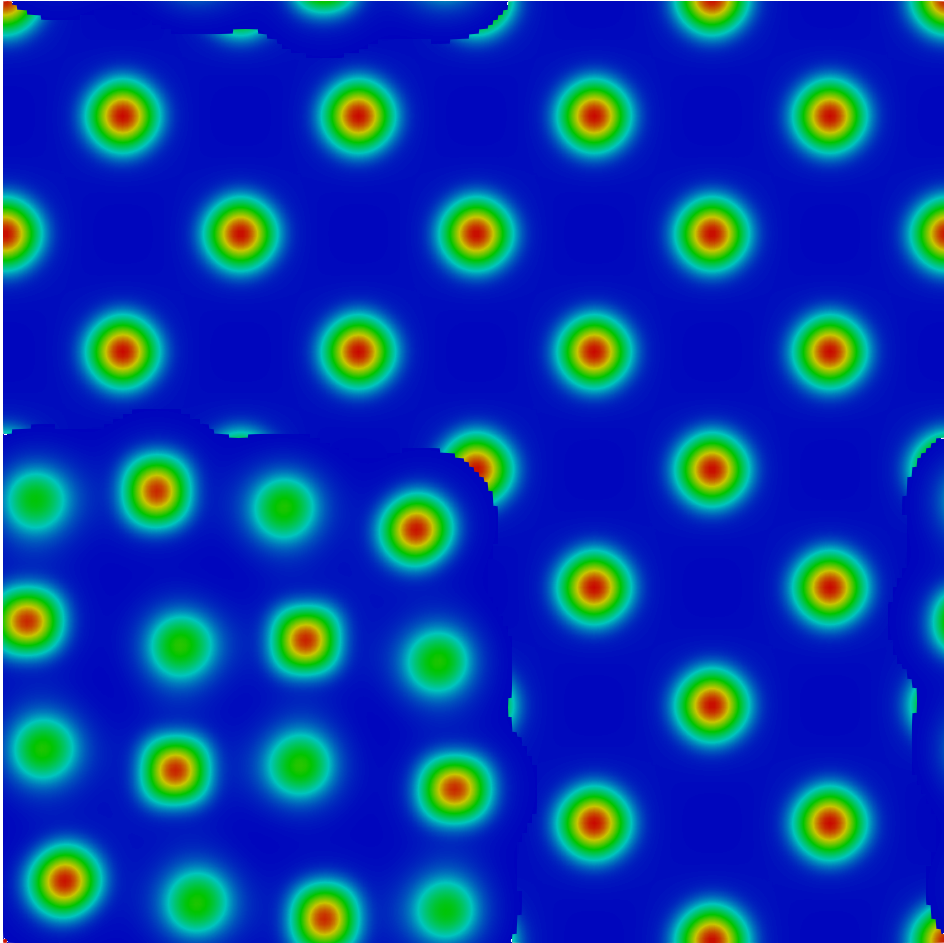}}
\caption{ Electrode charge density (top down view) in a region where a monolayer of NiO partially covers the Ni electrode (lower left). 
The cross section of the charge density of the partial NiO monolayer is superposed over the charge density of the underlying Ni surface.
}
\label{fig:electrode_charge_density}
\end{figure}

\begin{figure}[htp!]
\centering
{\includegraphics[width=1.0\figwidth]{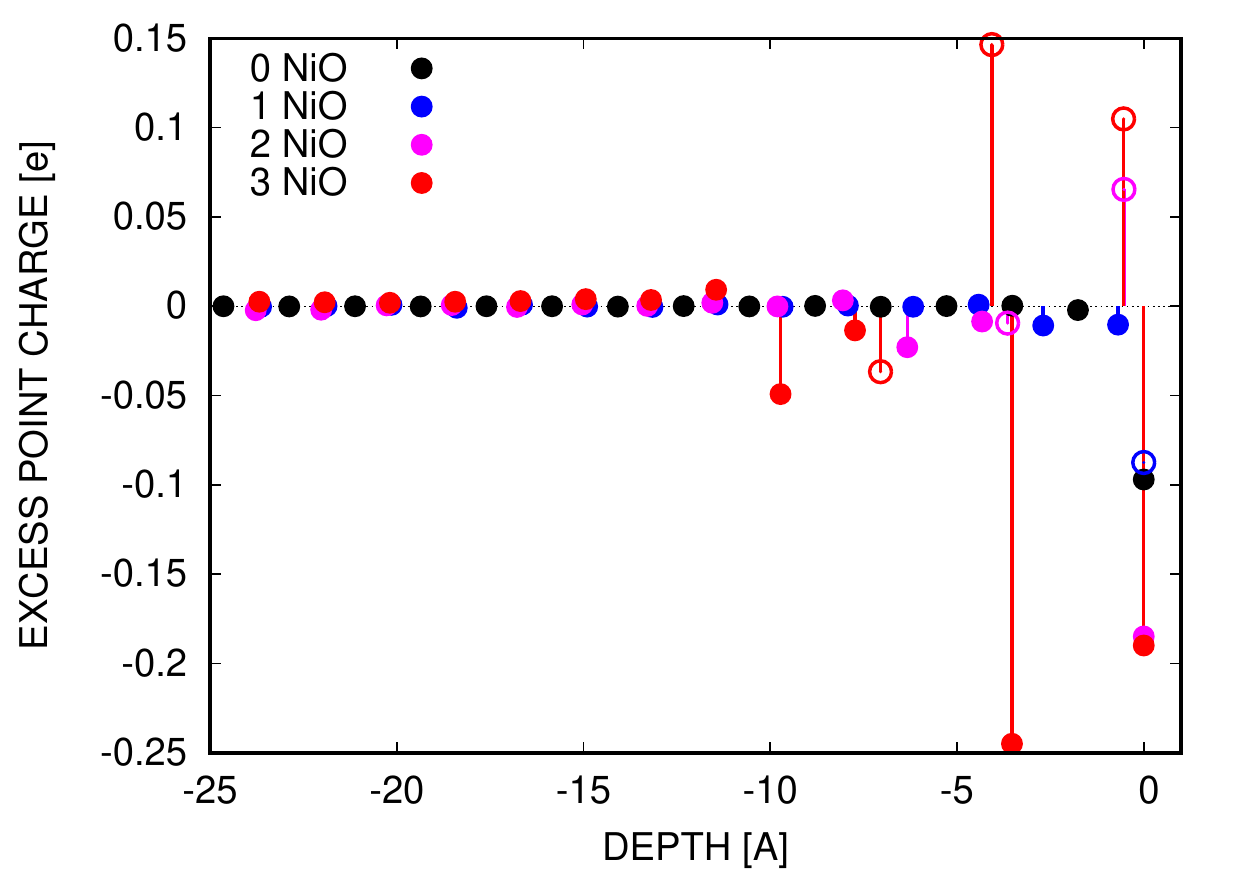}}
\caption{Excess point charges in the electrode: (black) bare Ni region, (color) region of NiO on Ni with 1:blue, 2:magenta, 3:red monolayers of NiO. Open circle marker:O, solid circle:Ni. Excess charge is relative to the expected valence, pure Ni:0, and Ni:+2 and O:-2 in the oxide layer. }
\label{fig:electrode_point_charges}
\end{figure}

Using first principles methods described in \sref{sec:electrode} we obtain the relaxed configuration and electron density of the electrode with and without an oxide layer.
\fref{fig:electrode_charge_density} shows a top down view of the charge density of a Ni surface partially covered by a NiO monolayer.
The NiO island is in the lower left of \fref{fig:electrode_charge_density} and the exposed portion of the underlying Ni surface can be seen in the remainder of the figure.
The initially square island relaxes to a more diamond-like shape but the variation in charge density across the partial layer is minimal, with charge transfer from the O to  Ni nuclei as expected.
Also, the charge density of the exposed Ni metal is relatively unperturbed by the oxide layer.
The mean separation between the Ni and the NiO layer is $\approx$2 \AA.  % 2.03 vs 2.35
With a sequence of related simulations we examined where the excess electrons (induced by electrode charging) reside.
\fref{fig:electrode_point_charges} plots the point charges (in excess of normal valence: metal Ni 0, oxide Ni +2, O -2) for bare Ni and Ni covered by 1, 2, and 3 monolayers of NiO
(the point charges for an overall neutral system and one with excess electrons are  extracted with a Bader method described in \sref{sec:method} and then differenced).
As expected, in the bare Ni, the excess electron density resides primarily in the outermost layer of Ni.
For a single layer of NiO covering Ni $\approx$ 80\% of the excess electrons reside near the O, $\approx$ 10\% near the Ni in the oxide layer and $\approx$ 10\% near the Ni in the underlying metal.
This charge splitting induces a surface dipole moment.
It is apparent that the charge distribution and the dipole effects become more complex as more oxide layers are added to the system.

\subsection{Bare metal region of the electrode}
With classical MD, we simulated the electrode-electrolyte systems over a sequence of surface charges: $\Delta q = \{ 0.0, 0.05, 0.10, 0.15\}$ $e$ per surface atom, corresponding to $\sigma = \{ 0.007, 0.014, 0.021, 0.028\}$ $e/\AA^2$.
These surface charges were related to voltage through the Gauss box technique described in \sref{sec:method}.

\begin{figure}[htp!]
\centering
{\includegraphics[width=1.2\figwidth]{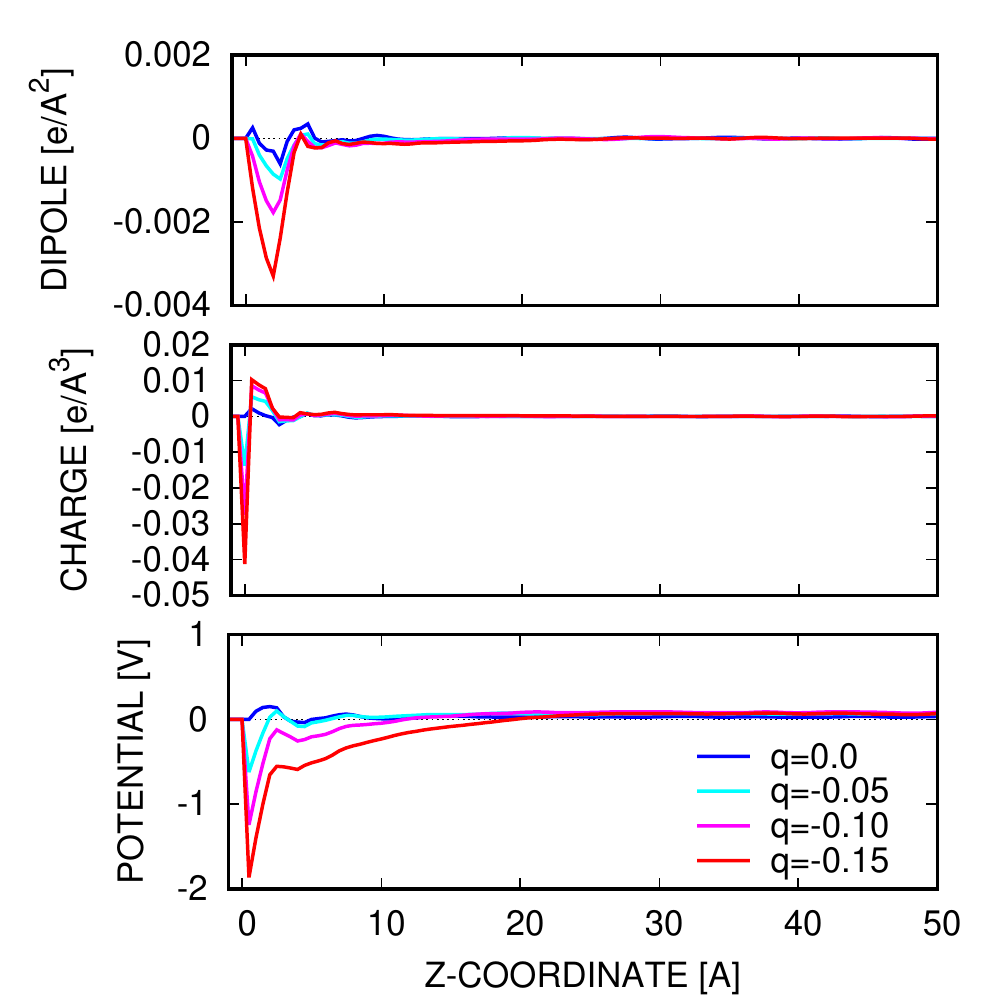}}
\caption{Bare Ni electrode: electrostatic potential, charge and water dipole density profiles as a function of surface charge $q$ [$e$/atom] on electrode.
}
\label{fig:potential-Ni}
\end{figure}

\fref{fig:potential-Ni} shows that the voltages at the electrode surface ($z$=0) range from 0 to $\approx$ 2 V for these surface charges.
Also, it is apparent from the charge density profiles that: (a) both a double layer next to the electrode and a diffuse layer grading to the bulk form exist, and (b) the decay to the bulk region with zero net charge occurs within 20\AA\ from the outer layer of the electrode.
We fit the potential profile in the diffuse region (approximately 5--20\AA\ from the electrode surface) to the Poisson-Boltzmann solution in \eref{eq:PB} and obtain $\ell \approx$ 0.5 \AA\ for $q =$ -0.1 $e$/atom.
This allowed us to estimate the effective dielectric in the diffuse layer $\epsilon = \frac{e \sigma}{2 k_B T} \ell \approx 24.4 \epsilon_0$ which is considerably less than pure water ($\epsilon \approx 80 \epsilon_0$).

In \fref{fig:potential-Ni} we also see that the region of significant dipole moment near the electrode decays to the bulk on the same scale as the potential that induces the orientation of the water molecules.
Most significantly, the dipole moment is mostly negative and increases in magnitude with applied bias, which corresponds to the expected conformation of the H atoms of the water molecules oriented closest to the negative electrode, and this orientation becoming more dominant with increased bias.
This finding is corroborated by direct observation, as in \fref{fig:surface_species} and \fref{fig:surface_orientation} (at  $q =$ -0.1 $e$/atom), where the white, partially positively charged H atoms lie on the electrode surface with the red, partially negatively charged O atoms of the same water molecules pointing away from the surface.
Either one or both hydrogens is directly associated with distinct metal surface atoms.
In a second layer, we observe some water molecules bridging water on the surface with their hydrogens associated with the oxygens of the surface adsorbed waters.
Even at zero bias it appears there is some orientational preference for the water molecules which must be induced by the particular LJ parameters for H and O atoms in the water molecules.
(The preferential alignment of water with uncharged surfaces is widely observed \cite{carrasco2012molecular}.)

\begin{figure}[htp!]
\centering
{\includegraphics[width=1.2\figwidth]{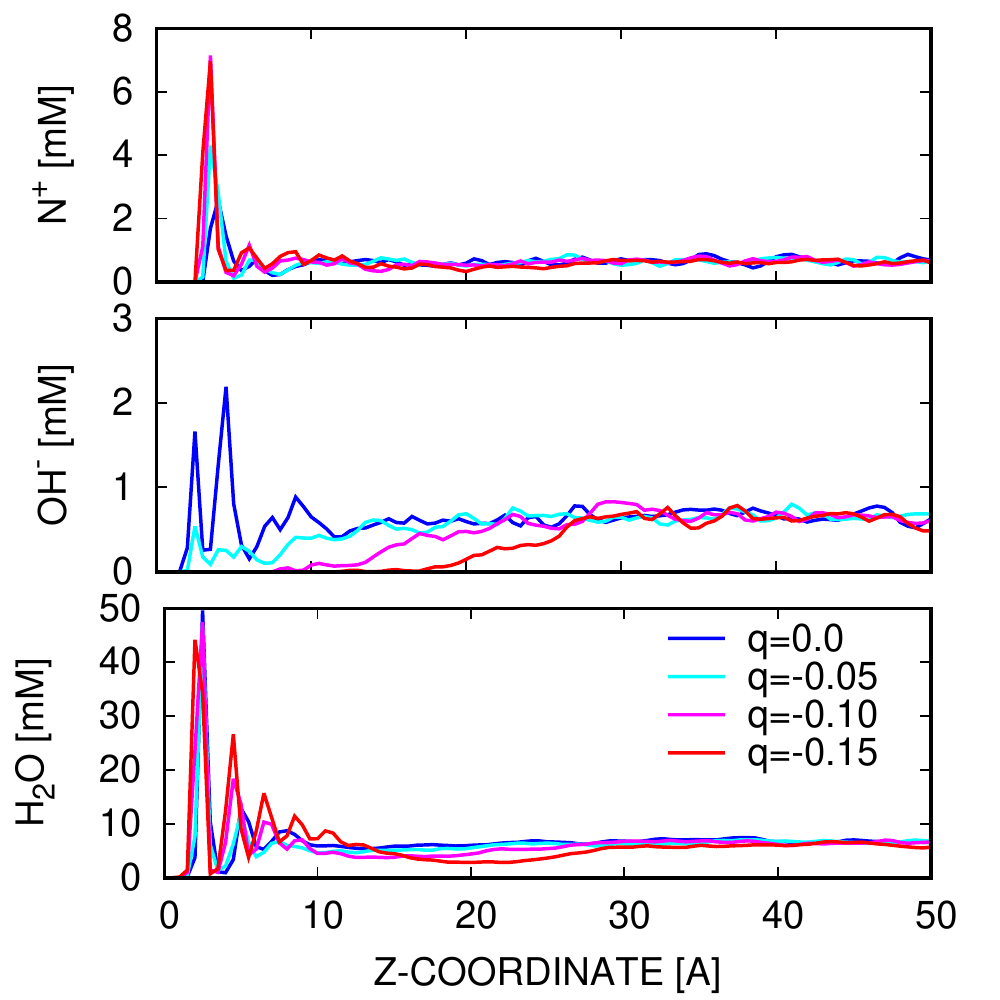}}

\caption{Bare Ni electrode: N$^+$, OH$^-$ and H$_2$O density profiles as a function of surface charge $q$ [$e$/atom] on electrode.
}
\label{fig:profiles-Ni}
\end{figure}

\fref{fig:profiles-Ni} shows that at zero bias, H$_2$O, OH$^-$ and the N$^+$ of the ionomer are adsorbed to the electrode surface and maintain uniform bulk concentrations away from the electrode. 
At higher biases OH$^-$ is excluded near the electrode, and water fills the vacated region and forms structured layers.
The spatial distribution of N$^+$ (averaged over time and the replica systems) is effectively uniform in the electrolyte region except there is a significantly higher concentration of N$^+$ at the interface (with corresponding neighboring region of depletion).
This appears to be due to the fact that ends of the chains are relatively mobile, unentangled and attracted to surface of the negative electrode.
The concentration of N$^+$ at the electrode more than doubles with the applied bias of $\approx$ 2 V and the response appears to be nonlinear, most likely due to an energy barrier in straightening the ends of the ionomer chains.
This can be seen as competition of elastic versus Coulomb forces for the relatively immobile N$^+$.
Note that these biases are high enough to dissociate water in the real system and the ionomer near the electrode might be likewise affected. 
The mobile species, H$_2$O and OH$^-$, respond as mentioned.
From \fref{fig:surface_species}, it apparent that no ring-like water structures  form and that, generally, the N$^+$ of the ionomer appear in regions of low concentrations of surface adsorbed water.
We presume that the near surface N$^+$ of the ionomer displace water and disrupt the tendency for the water to form regular structures.
The OH$^-$ transitions from significant surface adsorbed species at no bias to progressively vacated regions with increased voltage.
Here again, a threshold (determined by the relative magnitude of the overall electric field separating OH$^-$ and N$^+$ versus the LJ and Coulomb forces binding them) appears to be operating.
For biases less than $q=$-0.05 $e$/atom (not shown) OH$^-$ remains bound to N$^+$.
Note that effectively no OH$^-$ is present near the electrode at the $q=$-0.10 $e$/atom surface charge as shown in \fref{fig:electrode_charge_density}a.
Water, on the other hand, displays a surface adsorbed layer, a depletion zone, and then a uniform bulk concentration at no bias.
With increasing voltage more structured water layers appear near the surface and the closest layer becomes more packed (higher surface coverage).

\begin{figure}[htp!]
\centering
\begin{subfigure}[b]{0.45\textwidth}
{\includegraphics[width=1.0\figwidth]{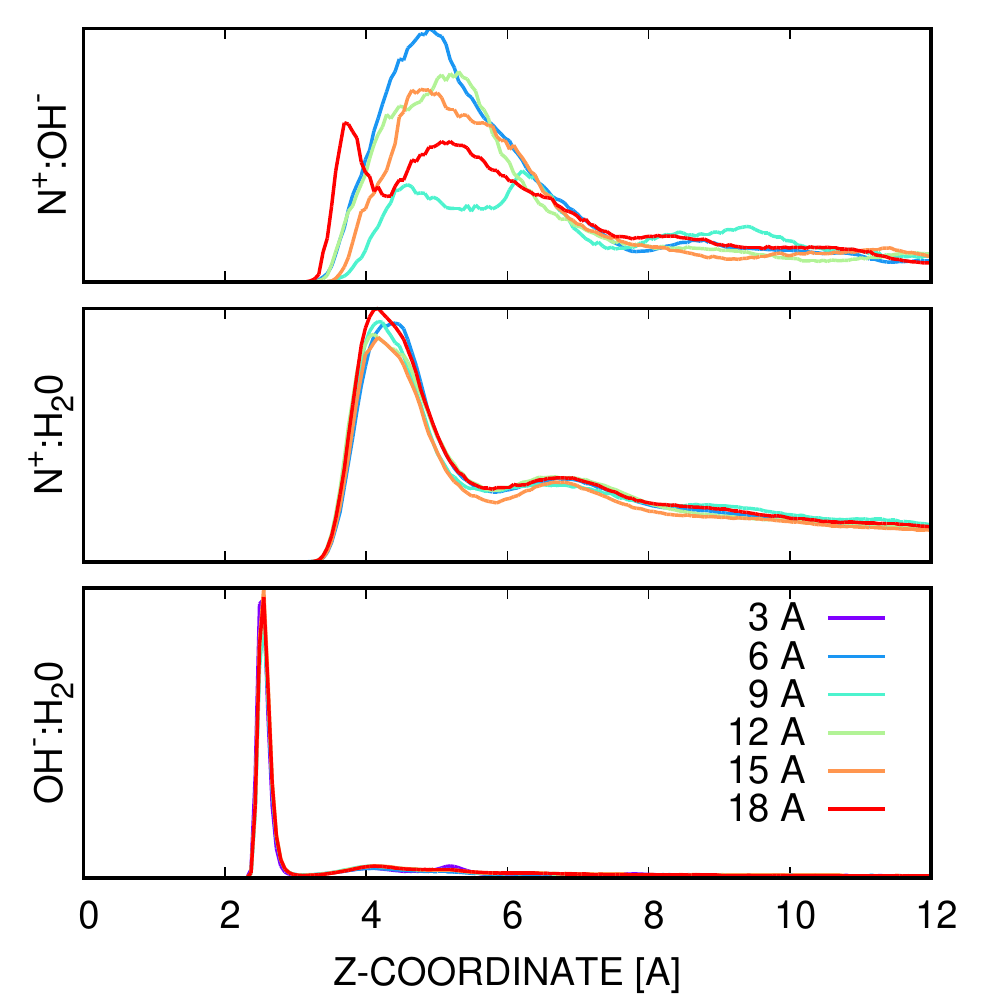}}
\caption{q=0.0}
\end{subfigure}
\begin{subfigure}[b]{0.45\textwidth}
{\includegraphics[width=1.0\figwidth]{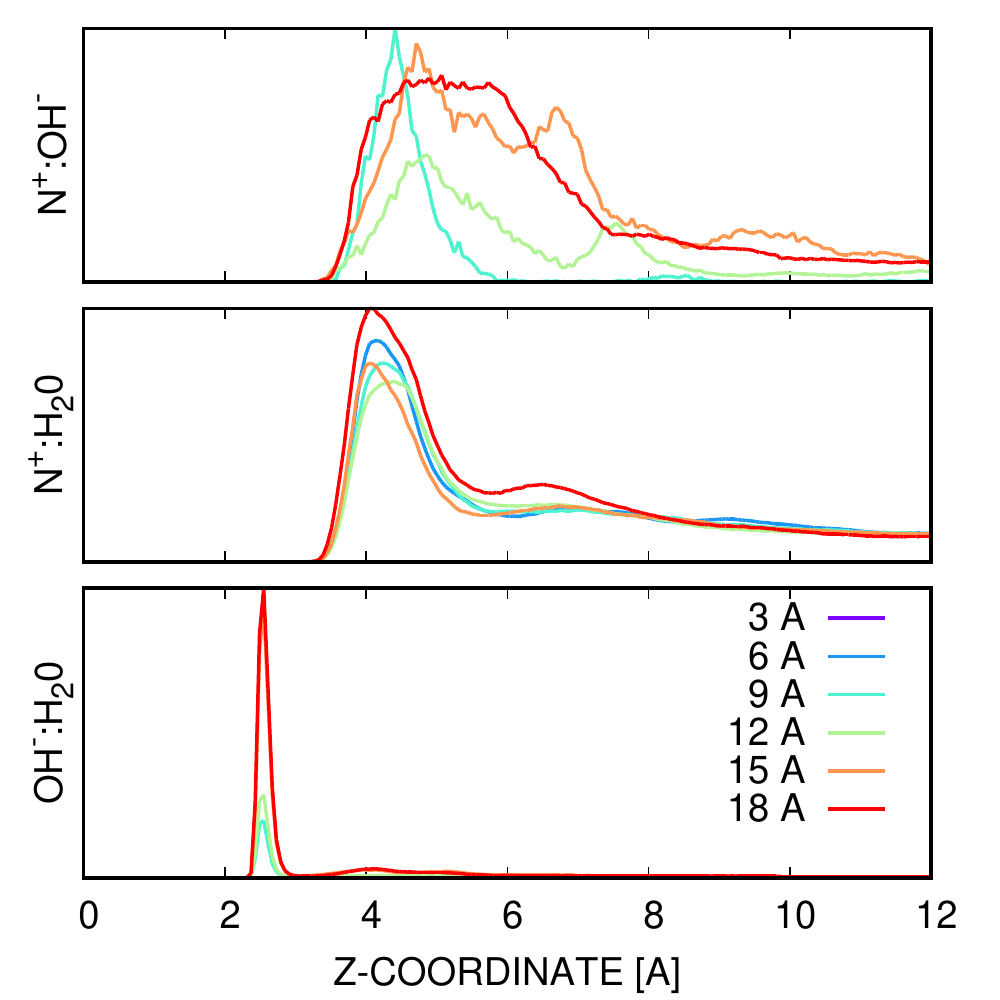}}
\caption{q=-0.1}
\end{subfigure}
\caption{Bare Ni electrode: radial distributions N$^+$:OH$^-$, N$^+$:H$_2$O, OH$^-$:H$_2$O as a function of distance from the electrode for surface charge (a) $q$=0.0 and (b) $q$=-0.1 [$e$/atom].
Radial distributions are computed for atoms in 3\AA\ wide regions starting at the electrode.
}
\label{fig:rdf_profiles-Ni}
\end{figure}

\fref{fig:rdf_profiles-Ni} shows the spatially binned RDFs for zero bias and $q=$-0.10 $e$/atom.
First of all, we see that OH$^-$ is closely coordinated with H$_2$O and this strong association does not change wherever OH$^-$ is present.
At voltage the only effect is decreasing the magnitude (not the location) of the RDF peak due to decreased OH$^-$ concentration near the electrode surface (refer to \fref{fig:profiles-Ni}).
Note no OH$^-$ is present at 0--6\AA\ from the electrode at $q=$-0.10 $e$/atom.
The coordination of N$^+$ with H$_2$O is uniform at zero bias and relatively unaffected at high bias which implies the OHP defined by H$_2$O solvation of N$^+$ is within 3 \AA\ of the outer layer of electrode surface.
The effects of distance from the electrode and bias on the coordination of N$^+$ with OH$^-$ are more complex due to changing local electric field and concentration.
Given the location of the RDF peaks, the primary coordination shell of OH$^-$ appears to mix with that of H$_2$O.
Most significant of the changes is that N$^+$ coordination with OH$^-$ tightens and sharpens at high bias and near the electrode, particularly in the 6--9\AA\ bin at $q$= -0.1 $e$/atom. 
In the next bin, 9--12\AA, we see two peaks corresponding to a two shell arrangement; and, by the 15--18\AA\ bin, the RDF resembles that of the bulk in both the zero and high bias cases.
If we use the coordination of N$^+$ with OH$^-$ to indicate the location of the OHP, we see the OHP shift away from the electrode surface reaching $\approx$ 20 \AA\ at the highest bias, as \fref{fig:profiles-Ni} indicates, due to the dissociation of the OH$^-$ from the N$^+$ at high bias.

\subsection{Oxide covered region of the electrode}
As observed in \fref{fig:surface_species} and \fref{fig:surface_orientation}, the presence of a (single layer of) NiO qualitatively changes the response of the HER electrode to electrical bias.
Comparing the surface coverage of a bare Ni region of the electrode with partially and fully covered NiO regions, we see approximately 40\% increase of adsorbed water where there is oxide on the surface.
\fref{fig:surface_species} also illustrates that there is distinctly increased water surface concentration at the NiO boundary.
The side view of the electrode in \fref{fig:surface_species} and the closeup in \fref{fig:surface_orientation} shows water molecules on the surface of the NiO covered electrode with one of the O-H legs of the water molecule flat on the surface and the other with H embedded in the oxide layer in direct association with an oxide O atom.
These effects are, in large part, due to the differences in the unbiased point charges and induced charge distribution (refer to \fref{fig:electrode_point_charges}).
Since oxides are dielectrics, they change local and long-range electric fields by screening due to induced and/or permanent dipoles.

\begin{figure}[htp!]
\centering
{\includegraphics[width=1.2\figwidth]{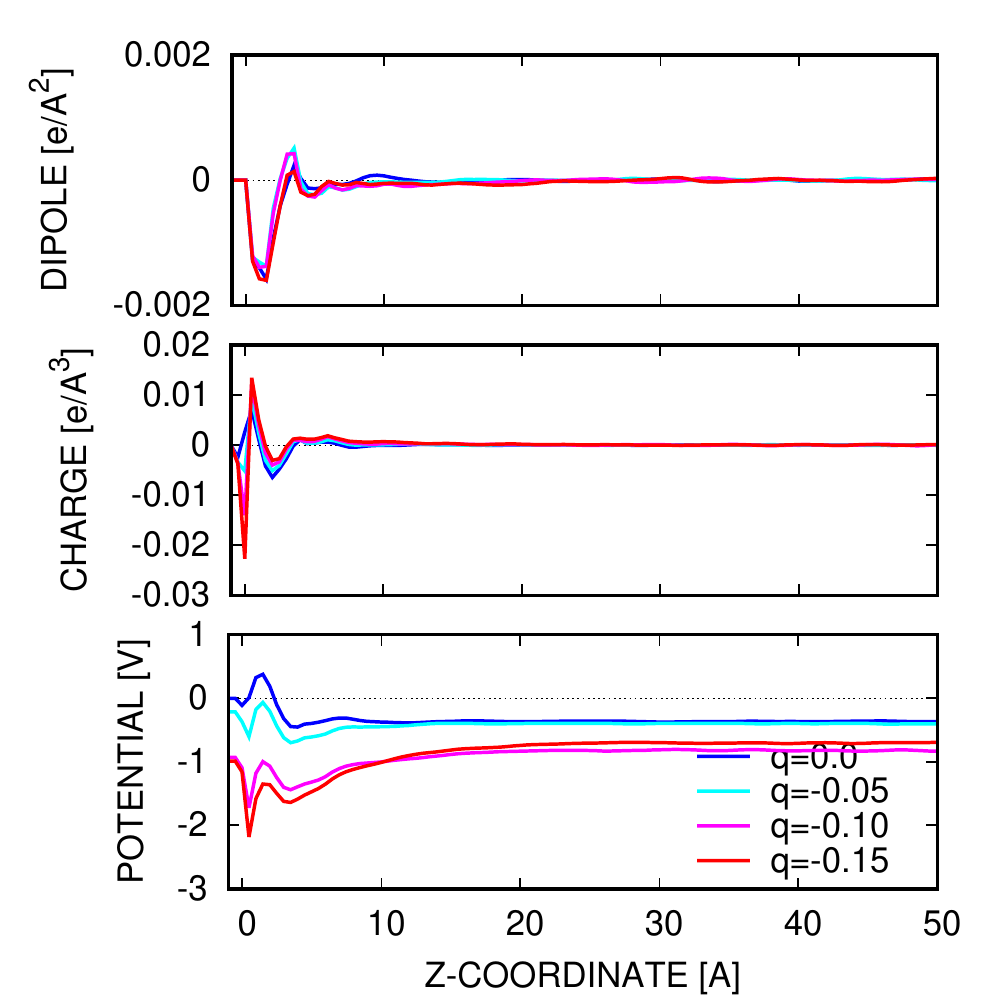}}
\caption{Electrode with NiO monolayer: electrostatic potential, charge and water dipole density profiles as a function of surface charge $q$ [$e$/atom] on electrode.
}
\label{fig:potential-NiNiO}
\end{figure}

At the same surface charges and comparable voltages to the simulation of the bare Ni region of the electrode discussed in the previous section, we compute the electrolyte's response to electrical bias for a Ni electrode with a monolayer of NiO as shown in \fref{fig:potential-NiNiO}.
There are some similarities and noticeable differences in the profiles compared to the bare Ni case shown in \fref{fig:potential-Ni}.
Firstly, there is a larger counter charge density next to the electrode in the NiO-on-Ni case, which we can associate with the partially positively charged H in the increased density of surface water molecules.
Examining the dipole density, the H-toward orientation is strong and barely shifts with increasing bias due to the strong, local polarization of the electrode itself.
This near-surface dipole density is reflected in the electric potential which becomes constant but not zero in the far-field.
This also leads to a lower effective voltage difference for the same surface charge, as expected with a dielectric layer.
In addition, the more prominent positive dipole peak seen in \fref{fig:potential-NiNiO} compared to \fref{fig:potential-Ni} is apparently due the dipole created by neighboring waters to since no inversions of water molecules with the H toward the surface were observed.

\begin{figure}[htp!]
\centering
{\includegraphics[width=1.2\figwidth]{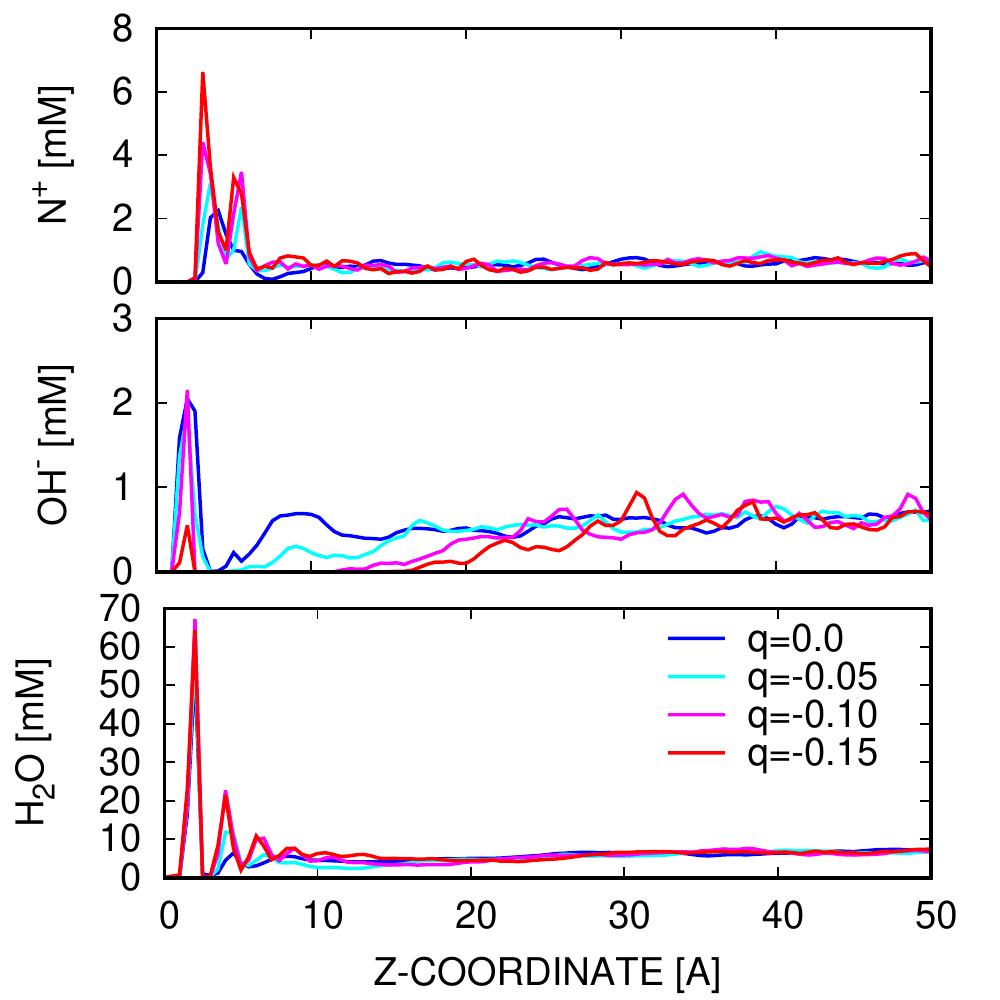}}

\caption{Ni electrode with NiO monolayer: N$^+$, OH$^-$ and H$_2$O density profiles as a function of surface charge $q$ [$e$/atom] on electrode.
}
\label{fig:profiles-NiNiO}
\end{figure}

Likewise the concentration profiles shown in \fref{fig:profiles-NiNiO} are comparable to \fref{fig:profiles-Ni} but also display distinct differences.
As in \fref{fig:profiles-Ni}, the water density is uniform in the far-field and shows distinct signs of structured layers near the electrode.
On the other-hand, the (time-averaged) water density at the electrode surface is much higher than in the bare metal case (approximately 40\%, as corroborated by the snapshots shown in \fref{fig:surface_species}).
Also, this spike in concentration is essentially unaffected by bias, a fact corroborated by nearly constant dipole density at the surface.
Remarkably, significant OH$^-$ remains associated with the electrode surface at high bias, but concentration decreases with increased bias (similar behavior has been observed in other systems \cite{lee2015atomistic}).
Otherwise, the OH$^-$ is similar to that for bare Ni, with a depletion zone increasing with increased bias and a nearly constant bulk concentration.
The N$^+$ concentration is also similar to that for the bare Ni, shown in \fref{fig:profiles-Ni}, but has a pronounced double peak near the interface.
This feature is perhaps related to significant residual OH$^-$ concentration at the surface with bias.

\begin{figure}[htp!]
\centering
\begin{subfigure}[b]{0.45\textwidth}
{\includegraphics[width=1.0\figwidth]{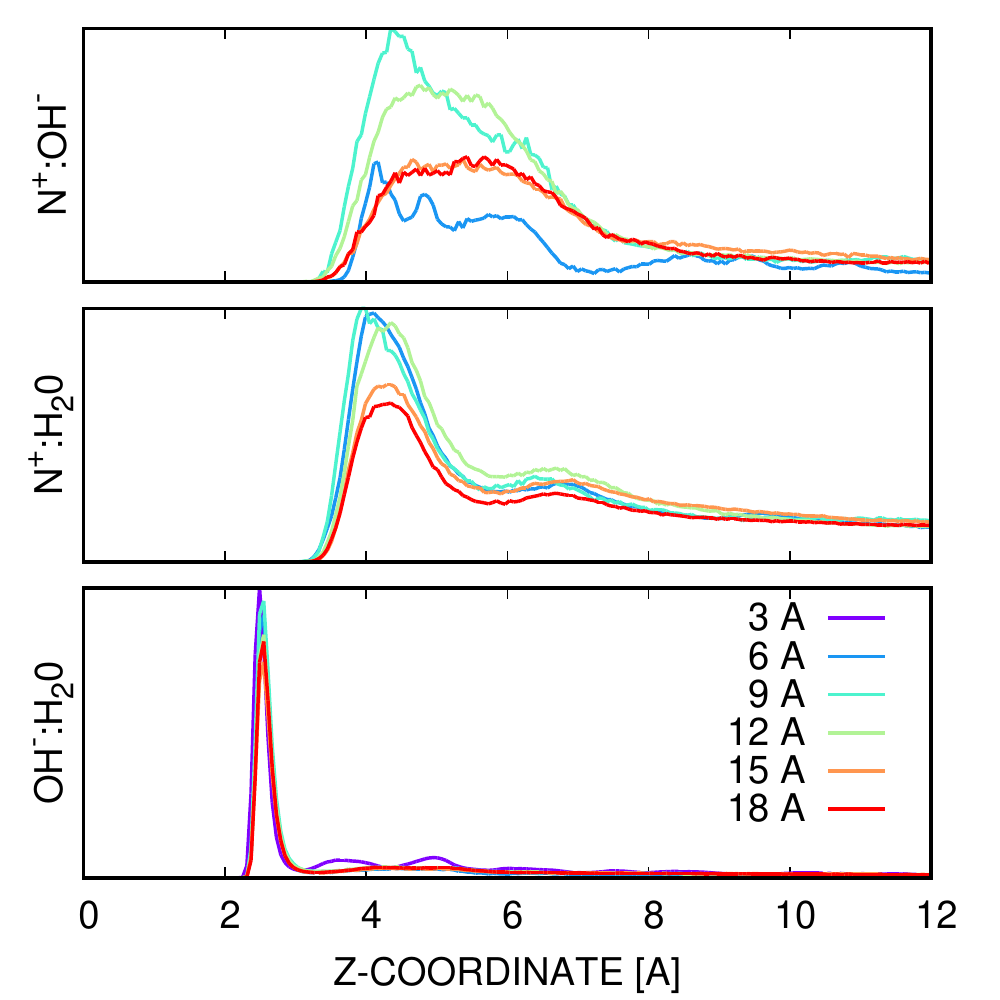}}
\caption{q=0.0}
\end{subfigure}
\begin{subfigure}[b]{0.45\textwidth}
{\includegraphics[width=1.0\figwidth]{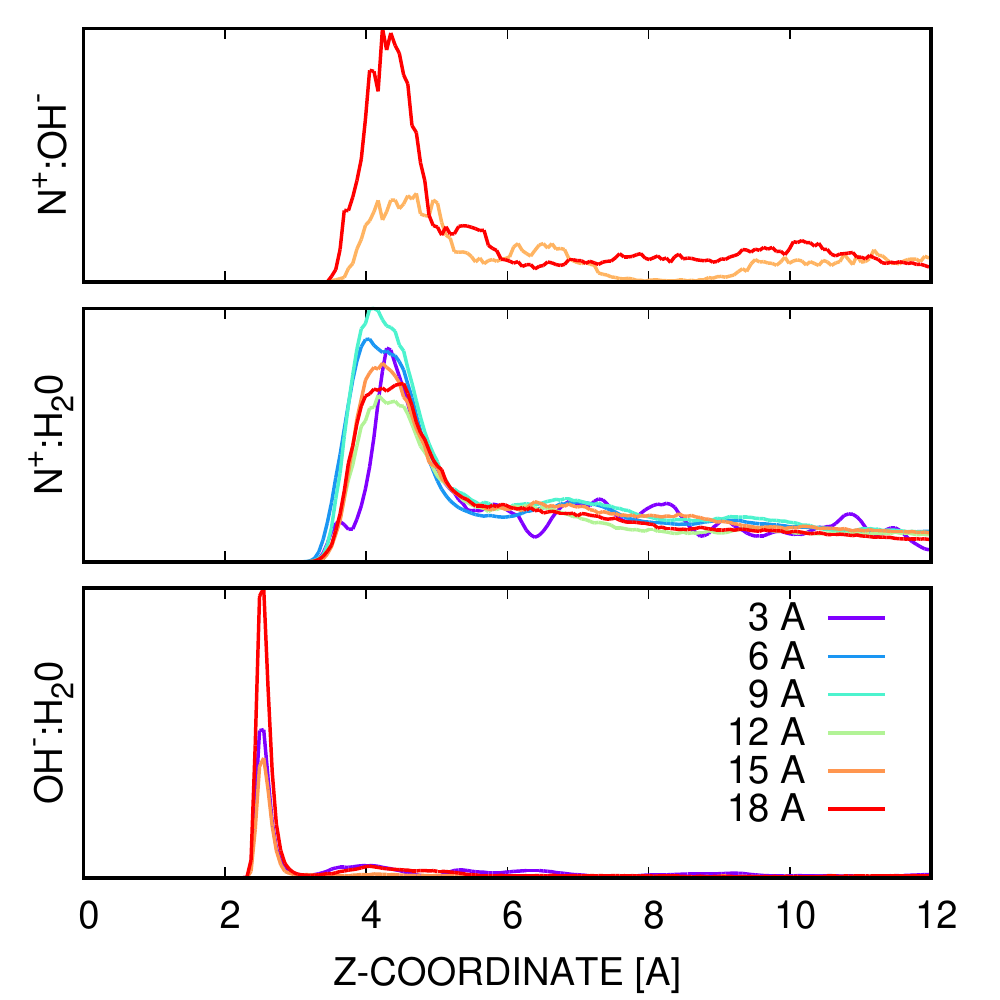}}
\caption{q=-0.1}
\end{subfigure}
\caption{Ni electrode with NiO monolayer: radial distributions N$^+$:OH$^-$, N$^+$:H$_2$O, OH$^-$:H$_2$O as a function of distance from the electrode for surface charge (a) $q$=0.0 and (b) $q$=-0.1 [$e$/atom].
Radial distributions are computed for atoms in 3\AA\ wide regions starting at the electrode.
Note the N$^+$:OH$^-$ distribution for N$^+$ in the 0--3\AA\ region is not plotted due to a noisy, low count average. 
}
\label{fig:rdf_profiles-NiNiO}
\end{figure}

The spatially binned RDFs shown in \fref{fig:rdf_profiles-NiNiO} are qualitatively similar to \fref{fig:rdf_profiles-Ni}, implying that the electrode surface is not radically changing the coordination structures.
There is a slight variation in OH$^-$ and H$_2$O concentration at low bias which affects OH$^-$:H$_2$O and N$^+$:H$_2$O coordination that is not observed in \fref{fig:rdf_profiles-Ni}.
The presence of OH$^-$ near the N$^+$ closest to the electrode in this case also appears in the N$^+$:OH$^-$ RDF but it is omitted in \fref{fig:rdf_profiles-Ni} due to sampling noise and an effort to maintain clarity.
Also, in this case the N$^+$:OH$^-$ RDF clearly relaxes to bulk distribution at zero bias.

\section{Discussion} \label{sec:discussion}

In response to the hypotheses posed in Bates \etal \cite{bates2015composite} and reiterated in the Introduction, we find:
\begin{itemize}
\item The electrical potential is altered by the ionomer and the oxide layer primarily due to dielectric effects of the immobile charges and their interplay with the mobile ones, as observed in \fref{fig:potential-Ni} and \fref{fig:potential-NiNiO}. 
A significant concentration of positive charge from the ionomer lies near the surface of the electrode, as can be inferred from \fref{fig:profiles-Ni} and \fref{fig:profiles-NiNiO}.
\item The ionomer N$^+$ are coordinated with both OH$^-$ and H$_2$O.
\fref{fig:rdf_profiles-Ni} and \fref{fig:rdf_profiles-NiNiO} indicate that solvation and hence the location of the OHP is changing with electrode bias voltage and depending on the solvating species ranges from 3 \AA\ off the electrode surface to 15 \AA.
A region depleted of OH$^-$ forms near the electrode with electrical bias and grows with increased bias.
Since the N$^+$ are relatively immobile, there are significant N$^+$ atoms in the depletion region, but these N$^+$ remain coordinated with water.
\item On the surface of the electrode, refer to \fref{fig:surface_species}, the N$^+$ in the ionomer apparently displaces water and disrupts the tendency of the water to form ordered surface structures like rings, as in \cref{nie2010pentagons}.
From \fref{fig:profiles-Ni} and \fref{fig:profiles-NiNiO} is it clear that significant concentration of the ionomer  N$^+$ lie near the electrode surface.
\item The water molecules near the electrode surface orient in the manner dictated by the electric field (with O further away from the surface) in a mixture of either one or both hydrogens associated with surface atoms; however, in the surfaces with an oxide layer the primary association is with a surface oxygen and the configuration where one of the H legs of the water molecule lies on the electrode surface seems to dominate, as can be seen in \fref{fig:surface_species} and \fref{fig:surface_orientation}.
\item Approximately 40\%  of the bare metal surface is coordinated with  water molecules and this surface coverage almost doubles in areas covered by an oxide layer.
\item Nanoscale heterogeneity such as at the edge of a oxide region provides local configurations favorable for water adsorption, as evidenced by the regions of high water coverage at the edge of the partially oxidized electrode in \fref{fig:surface_species} and \fref{fig:surface_orientation}.
This qualitative observation may corroborate how rough versus smooth electrodes exhibit qualitatively different reactivity, see \cref{thiel1987interaction} (Sec. 3.1 and 4.2) and references therein.
\item The presence of a NiO monolayer apparently allows OH$^-$ to remain near the electrode at high negative surface charges, compare \fref{fig:profiles-Ni} and \fref{fig:profiles-NiNiO}.
\end{itemize}
Although we are presently unable to simulate the relevant reactions at the necessary time- (molecular diffusion) and length- (diffuse layer) scales, the fact that the reactions are fast compared with the processes MD represents well gives us some confidence that the results are representative of steady-state conditions. 
Also, the H$_2$ produced at the HER electrode is  small and mobile, and not likely to affect the components of the electrolyte.
If positive ions are present \eg K$^+$ from dissolved KOH, they will likely absorb to the electrode surface and hence screen the remainder of the electrolyte, having stronger effects on performance.

\section{Conclusion} \label{sec:conclusion}

We were able to simulate and examine the HER electrode-ionomer electrolyte interface with atomic detail using a combination of DFT and classical MD techniques.
We observed configuration changes in response to external bias and the oxide coverage of the electrode.
Information of this nature is relevant to the efficiency of the water splitting process. 
In particular the concentration of the reactants, water, and the electric field at the interface have a direct relation to H$_2$ production.

In future work, we will explore in more detail the effect of surface roughness/nano-structuring and the presence of other phases, such as Cr$_2$O$_3$, on the HER electrode.
Resolution of the dielectric field \cite{mandadapu2013polarization} may also shed light on the operation of these types of water splitting cells and give rise to more accurate and informative theories.
Also, the entropic changes due to changes in pH discussed in \cref{rossmeisl2016ph} are likely correlated with structural characterizations presented in this work, such as the spatial variation of radial distributions, and are another topic for potential future work.

\section*{Acknowledgments}
We would like to thank Norman Bartelt and Jeremy Templeton (Sandia) for insightful discussions on this work.
This work enabled by VASP (TU Vienna, \verb"https://www.vasp.at"), 
LAMMPS (Sandia, \verb"https://lammps.sandia.gov"), and 
Bader (Univ. Texas \verb"http://theory.cm.utexas.edu/henkelman/code/bader/").
This work was supported by the U.S. Department of Energy, Office of Energy Efficiency and Renewable Energy (EERE), specifically the Fuel Cell Technologies Office
Sandia National Laboratories is a multimission laboratory managed and operated by National Technology and Engineering Solutions of Sandia, LLC., a wholly owned subsidiary of Honeywell International, Inc., for the U.S. Department of Energy's National Nuclear Security Administration under contract DE-NA0003525.
The views expressed in the article do not necessarily represent the views of the U.S. Department of Energy or the United States Government.

%%%%%%%%%%%%%%%%%%%%%%%%%%%%%%%%%%%%%%%%%%%%%%%%%%%%%%%%%%%%%%%%%%%%%%%%
%
%%%%%%%%%%%%%%%%%%%%%%%%%%%%%%%%%%%%%%%%%%%%%%%%%%%%%%%%%%%%%%%%%%%%%%%%
\end{document}